\documentclass[prl,aps,onecolumn,showpacs,preprintnumbers,amsmath,amssymb,
floatfix]{revtex4-1}
\usepackage{enumerate}
\usepackage{graphicx}
\usepackage{bm}
\usepackage{amsmath, amsthm, amssymb}

\theoremstyle{remark}

\usepackage{enumerate}
\usepackage{graphicx}
\usepackage{amsmath, amssymb}
\usepackage{graphics}
\usepackage{bm}
\begin{document}
\newcommand{\real}{\textrm{Re}\:}
\newcommand{\sto}{\stackrel{s}{\to}}
\newcommand{\Tr}{\textrm{Tr}\:}
\newcommand{\Ran}{\textrm{Ran}}
\newcommand{\supp}{\textrm{supp}\:}
\newcommand{\evs}{\textnormal{evs}\:}
\newcommand{\Ker}{\textnormal{Ker}\:}
\newcommand{\sign}{\textnormal{sign}\:}
\newcommand{\Resi}{\textnormal{Res}\:}
\newcommand{\wto}{\stackrel{w}{\to}}
\newcommand{\ssto}{\stackrel{s}{\to}}
\newcommand{\epstr}{{\tilde \epsilon}_0}
\newcommand{\beps}{\pmb{\epsilon}}
\newcounter{foo}
  \newcommand{\sslim}{\textnormal{s--}\lim}
\newcommand{\wlim}{\textnormal{w--}\lim}
\providecommand{\norm}[1]{\lVert#1\rVert}
\providecommand{\abs}[1]{\lvert#1\rvert}
\providecommand{\absbm}[1]{\pmb{|}\mspace{1mu}#1\mspace{1mu}\pmb{|}}

\title{Manifestation of Universality in the Asymmetric Helium Trimer and in the Halo Nucleus $^{22}$C}

\author{Dmitry K. Gridnev}
\affiliation{FIAS, Ruth-Moufang-Stra{\ss}e 1, D--60438 Frankfurt am Main,
Germany}

\author{Dario Bressanini}
\affiliation{Universita dell'Insubria, Via Valleggio 9, 22100 Como Italy}

\begin{abstract}
We prove that the corner angle distributions in the bound three-body system AAB, which consists of two particles of type A and one particle of type B, approach universal form if the pair AA has a virtual state at zero energy and the binding energy of AAB goes to zero. We derive 
explicit expressions for the universal corner angle distributions in terms of elementary functions, which depend solely on the mass ratio m(A)/m(B) and do not depend on pair interactions. On the basis of experimental data and calculations we demonstrate that such systems as the asymmetric Helium trimer $^3$He$^4$He$_2$ and the halo nucleus $^{22}$C exhibit universal features. Thus our result establishes an interesting link between atomic and nuclear physics through the few-body universality. 
\end{abstract}

\maketitle

%%%%%
%%%%%                       MAIN   PART      %%%%%%%%%
%%%%%%%%%%%%%%%%%%%%%%%%%%%%%%%%%%%%%%%%%%%%%%%%

%\section{Introduction}

The last ten years saw an enormous progress in the field of Efimov 
physics, theoretically as well as experimentally \cite{1,2,4,5,6,7,8,9,10,11,12,13}. Back in 1970 V. Efimov predicted \cite{14} the existence
of universal three-body bound states with a geometric spectrum for identical bosons at the infinite scattering length. Efimov's
counterintuitive prediction was that just by tuning the strength of short-range interactions in the 3-body system one can bind an
infinite number of levels even though the two-body subsystems remain unbound. The wave functions even of the lowest Efimov states are very diffuse and exhibit an enormous spatial extension, which by far exceeds the scales of the underlying short range interactions. The energies of the levels are universally related, the ratio 
of the adjacent energy levels $E_n$ and $E_{n+1}$ quickly approaches the formula $E_n /E_{n+1} = e^{2\pi /s_0}$, where $s_0 \simeq 1.00624$ is a universal constant \cite{14,1}. Universal here means that this relation does not depend on the particular form of pair interactions in the three-body system. 

The term universality refers to the fact that physical systems that are completely different on short scales can in certain limits exhibit identical behavior. A well-known example is a universal value of certain critical exponents corresponding to phase transitions near the critical points \cite{1}. This value happens to be identical for the substances that are very different on the microscopic scale. Universal behavior originates from the long-range 
order in the system, which arises at the critical point and makes the details of the pair interaction irrelevant. Universality in few-body systems is now being extensively explored \cite{1}, universal features have been found in the structure of the wave function \cite{my3} and in lower dimensional systems \cite{nishida,my4}. 

The Efimov state was first registered experimentally in the ultracold gas of Caesium atoms \cite{4}. The trapped gas was placed into a magnetic field and the Feshbach tuning allowed the resonant formation of Efimov trimers. The similar technique allowed to find a second Efimov state \cite{4.5} 
and verify the Efimov's prediction regarding the universal scaling. 
These states were observed indirectly through a giant three-body recombination loss appearing at the certain values of the magnetic field. 
Through measuring the enhancement of recombination the Efimov effect has by now been observed in
bosonic isotopes of potassium \cite{6}, and lithium \cite{13}. Unfortunately this technique does not allow any insight into the inner structure of the trimers.

A recent experiment \cite{doerner1}, where a long predicted bound state of the Helium trimer was detected, 
marked a new milestone in the search of Efimov states. Helium trimer is a naturally existing molecule 
consisting of 3 very weakly bound $^4$He atoms. In the experiment cold  $^4$He atoms were released from a 5 $\mu$m nozzle onto the grating 
with spacing 50 $\mu$m, which resulted in the diffraction pattern. At the point, 
where kinematically one could expect the clustering of trimers, the fast laser pulse 
stripped off the electrons. In such event the molecule was adiabatically translated the into the 
position, where naked nuclei would be at the starting point of the process called the "Coulomb explosion". The 
detectors captured the outgoing particles and the geometrical structure 
of the trimer corresponding to the moment just before the laser pulse could be 
reconstructed. After processing many such events one can plot, in particular, 
distributions of corner angles in the molecule \cite{doerner1}. The asymmetric Helium trimer $^3$He$^4$He$_2$, which is the helium trimer with one atom being replaced by the lighter isotope $^3$He, has been observed in a similar experiment \cite{doerner2}. These weakly bound 
systems possess large spatial extensions and are quantum halos, yet the Efimov 
universality could not be found because the second Efimov state in the trimer $^4$He$_3$ is 
unstable and cannot be detected. The aim of the present letter is to show that the universality manifests itself in the corner angle distribution of the asymmetric Helium trimer. We derive explicit expressions in elementary functions for the universal corner angle distributions, which depend only on the mass ratios and thus are independent of the form of pair interaction. We demonstrate that the observed corner angle distributions in the asymmetric Helium trimer match to a large extent the universal ones. Hopefully, in future experiments one could combine the laser pulse ionization technique \cite{doerner1,doerner2} with traps \cite{4,4.5} so that one would get an insight into the internal structures of trimers other than Helium. 

Although experimentally verified only in molecules the original Efimov's prediction \cite{14} actually concerned nuclear systems. However, the experimental search of 
this effect in nuclei is impeded by the fact that the nuclear forces between nucleons and nuclear clusters cannot be easily manipulated. 
One can only count on an ``accidental'' tuning \cite{1} in the sense that the interaction between particles (clusters) is resonant. The promising candidates, where universality can be looked for, are halo nuclei. Halo nuclei \cite{vaagenreports} are very weakly-bound exotic isotopes in which the outer two valence nucleons are spatially decoupled from a tightly bound core such that they 
locate dominantly in the classically forbidden region. Halo nucleons tunnel out to large distances giving rise to extended wave function tails and hence large overall matter radii. The carbon isotope $^{22}$C is believed to be the nucleus having the largest so far detected halo formed by two neutrons orbiting around the core $^{20}$C \cite{carbonexp,vaagencarbon,horiuchi,tomiocarbon}. Here we would show that the nucleus $^{22}$C viewed as a 3-body system consisting of the core and two neutrons exhibits universal features reflected in its corner angle distributions.  
By that we demonstrate the true power of universality, which establishes a link between seemingly unrelated atomic and nuclear systems, namely the 
asymmetric helium trimer and the halo nucleus $^{22}$C. 

\begin{figure}
\includegraphics[height=0.16\textheight]{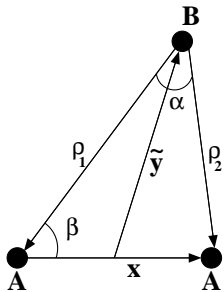}
\caption{\label{Fig:coord} Various coordinates in the 3-body system and the corner angles. Here ${\pmb x} = {\pmb \rho}_2 -{\pmb \rho}_1$ and $\tilde {\pmb y} = -(1/2)({\pmb \rho}_1 + {\pmb \rho}_2)$. }
\end{figure}

Consider a 3-body system consisting of 2 identical particles A with mass $m_A$ and one 
particle B with mass $m_B$ (particles A can be either bosons or fermions sitting in 
different spin states). Let the particles interact 
through short-range potentials $V_{AA} (| {\pmb \rho}_1 - {\pmb \rho}_2  |)$ and $V_{AB} (|{\pmb \rho}_1|)$ (the 
vectors ${\pmb \rho}_{1,2}$ are illustrated in Fig.~\ref{Fig:coord}). The Hamiltonian of the 
system reads 
\begin{equation}\label{eq:1}
H = T + \lambda_{AA} V_{AA} (| {\pmb \rho}_1 - {\pmb \rho}_2  |) + \lambda_{AB} \sum_{i=1}^2 V_{AB} (|{\pmb \rho}_i|) 
, 
\end{equation}
where $T$ is the kinetic energy operator with the removed center of mass motion 
and $\lambda_{AA}, \lambda_{AB} >0$ are the coupling constants. Since we 
have introduced the coupling constants we can assume without loosing generality 
that two particles of type $A$ interacting through the potential $V_{AA}$ have a 
virtual state exactly at zero 
energy but no bound states with negative energy (this leads to an infinite 
scattering length). Similarly we assume that the particles $A$ and $B$ interacting through the potential $V_{AB}$ 
also have a virtual state exactly at zero energy but no bound states with negative energy.

\begin{figure}
\includegraphics[height=0.3\textheight,angle=-90]{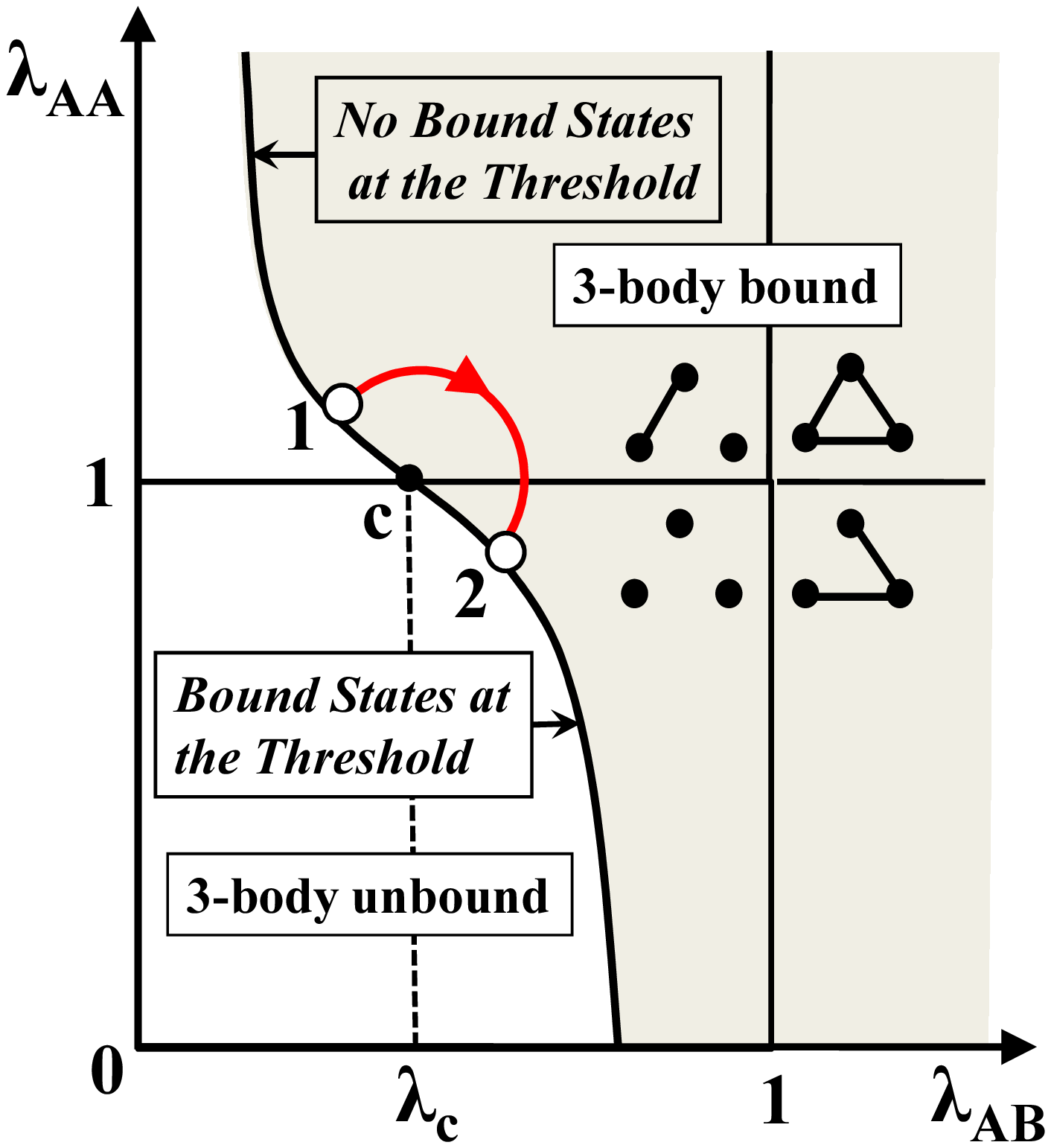}
\caption{\label{Fig:diagram}(Color online). The sketch of the typical stability diagram of a 3-body system in terms of couplings. The shaded area corresponds to stable systems, which have at least one bound state lying below the dissociation spectrum. The stability curve separates stable and unstable areas. 
A path connects the stable points lying very close the stability curve, whereby points 1 and 2 lie respectively above and below the critical point. The pictograms indicate the regions of Borromean, ```tango'', ``samba'' and ``all-bound'' configurations. The change in the corner angle distribution along the path is illustrated in Fig.~\ref{Fig:bifurcation}.}
\end{figure}

The system of 3 particles $AAB$ 
described 
by the Hamiltonian Eq.~(\ref{eq:1}) is {\bf stable} 
if it has a bound state with the energy lying below the 3-body continuum. The 
stability diagram of 3 particles $AAB$ in terms of coupling constants \cite{hansen,richard} is schematically 
illustrated in Fig.~\ref{Fig:diagram}, where stable and unstable systems 
are separated by the {\bf stability curve}. In the square, where $0 < \lambda_{AA} 
\leq 1$ and $0 < \lambda_{AB} \leq 1$, the stable systems are Borromean 
\cite{vaagenreports}, which means they do not have 
bound subsystems. The area, where $\lambda_{AA} > 1$ and $0 < \lambda_{AB} \leq 1$, 
represents the ``tango'' configuration \cite{garrido,yamashita}, when we have two unbound and one bound two-body subsystems. The area 
$0 < \lambda_{AA} 
\leq 1$ and $ \lambda_{AB} > 1$ corresponds to the ``samba'' configuration, when just one two-body subsystem is unbound; the rest of the diagram are 
the ``all-bound'' configurations, where every particle pair has a bound state. The sectors of all these configurations are denoted with pictograms in Fig.~\ref{Fig:diagram}. The line $\lambda_{AA}=1$ crosses the stability curve in the so-called {\bf critical point}. 
Below we discuss what happens to the ground state wave function of the 3-body system when the point on the stability diagram approaches the stability 
curve from the right (in the shaded area in the diagram Fig.~\ref{Fig:diagram} all systems have a well-defined normalized ground state wave function).

\begin{figure}
\includegraphics[height=0.25\textheight]{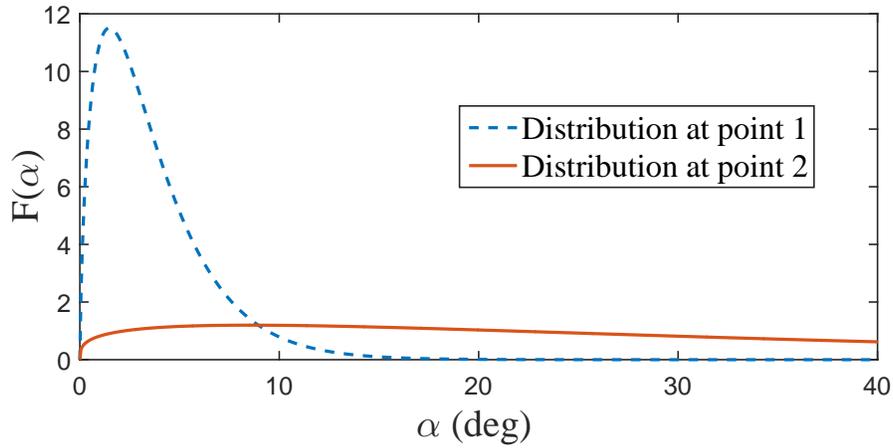}
\caption{\label{Fig:bifurcation}(Color online). Drastic change in the corner angle distribution $F(\alpha)$ when a point on the stability diagram 
travels along the path shown in Fig.~\ref{Fig:diagram}. The path leading to such transition can be arbitrarily short.}
\end{figure}

We shall use the normalized Jacobi coordinates ${\pmb x} = \hbar^{-1}\sqrt{m_A}({\pmb r}_2 -{\pmb r}_1)$ and ${\pmb y} = \hbar^{-1}\sqrt{m_A m_B /(2m_A +m_B)}(2{\pmb r}_3 - {\pmb r}_1 - {\pmb r}_2)$, where ${\pmb r}_i$ are position vectors and particles $\{1,2\}$ are of type A, particle 3 is of type B. We shall always assume that $\int |\psi({\pmb x},{\pmb y})|^2d{\pmb x}d{\pmb y} =1$, where $\psi({\pmb x},{\pmb y})$ is the bound state wave function. 
On the stability curve the ground state energy equals the energy at the bottom of the continuum and regarding the behavior of energies and wave functions there are 3 possible scenarios \cite{my5}. 
If the system approaches the point on the stability curve, where $\lambda_{AA} < 1$, its 
wave function does not spread \cite{my1,my2,my3}, the particles remain confined and for a point lying exactly 
on the stability curve there exists a well-defined 
normalized ground state wave function, which corresponds to zero energy. This exotic zero energy ground state wave function 
does not decay exponentially but rather falls off like an inverse polynomial \cite{my1,my3}. 
The second scenario is realized when the system approaches 
the stability curve, at the point where $\lambda_{AA} > 1$. In that case the 3-body wave function $\psi({\pmb x},{\pmb y})$ totally spreads \cite{my1} and approaches a universal expression \cite{klaus,my3}, namely,  
\begin{equation}\label{eq:3}
 \lim_{E\to 0}\left\| \psi ({\pmb x},{\pmb y}) -  \frac 1{4\pi} \phi_{AA} ({\pmb x}) \frac{e^{-\sqrt{|E|}|{\pmb y}|}}{|{\pmb y}|}\right\| = 0 . 
\end{equation}
In Eq.~(\ref{eq:3}) $E$ is the binding energy (energy of the bound state minus the energy of the lowest dissociation threshold) and $\phi_{AA} ({\pmb x})$ is the wave function of the bound pair $AA$ at the point, where the stability curve is hit. Suppose that the line $\lambda_{AA} = \textnormal{const} > 1$ crosses the stability curve in the point where $\lambda_{AB} = \lambda_{AB}^0$. Let us denote by $E(\lambda_{AB})$ the energy of the stable system lying on the line $\lambda_{AA} = \textnormal{const}$. Then there is a constant c such that \cite{my5,klaus} 
\begin{equation}
\lim_{\lambda_{AB}\to \lambda_{AB}^0} c E(\lambda_{AB})(\lambda_{AB} - \lambda_{AB}^0)^{-2} =1
\end{equation}

The third scenario occurs when the stable system approaches the critical point along the line $\lambda_{AA} =1$. In this case \cite{my3}
\begin{equation}\label{eq:4}
 \lim_{E\to 0} \left\| \psi({\pmb x},{\pmb y}) -  \frac{\chi_{[1,\infty)}(\rho)}{\pi^{3/2} |\ln
|E||^{1/2}} \frac{ \bigl\{ |{\pmb x}|\sin(\sqrt{|E|} |{\pmb y}|) + |{\pmb y}|\cos(\sqrt{|E|} |{\pmb y}|)\bigr\}e^{-\sqrt{|E|}|{\pmb x}|}
}{|{\pmb x}|^3|{\pmb y}|+|{\pmb y}|^3 |{\pmb x}|} \right\| = 0 . 
\end{equation}
In Eq.~(\ref{eq:4}) $\rho^2 = {\pmb x}^2 + {\pmb y}^2 $ and by definition $\chi_{[1,\infty)}(\rho) = 1$ for $\rho \in [1, \infty)$ and $\chi_{[1,\infty)}(\rho) = 0$ otherwise.

Let us denote by $E(\lambda_{AB})$ the energy of the stable system lying on the line $\lambda_{AA} = 1$. Then there exists such constant $c$ that \cite{my5}
\begin{equation}\label{eq:5}
\lim_{\lambda_{AB}\to \lambda_{cr}} c E(\lambda_{AB})\bigl[ \ln (\lambda_{AB} - \lambda_{cr})\bigr] (\lambda_{AB} - \lambda_{cr})^{-1} =1 ,  
\end{equation}
where $\lambda_{cr}$ is the value of $\lambda_{AB}$ at the critical point. The same equation can be rewritten in terms of the scattering length 
\begin{equation}\label{eq:6}
\lim_{a\to a_{cr}} c' E(a)\bigl[ \ln |a - a_{cr}|\bigr] (a - a_{cr})^{-1} =1 ,  
\end{equation}
where $a$ is the scattering length for the pair of pair of particles $AB$ interacting through $\lambda_{AB} V_{AB}$ and $a_{cr}$ is its value at the critical point and $c'$ is another constant. Eq.~(\ref{eq:6}) follows from Eq.~(\ref{eq:5}) because $a_{cr}$ is finite and $a$ can be Taylor expanded in terms of the coupling constant. Eqs.~(\ref{eq:5})--(\ref{eq:6}) and Eqs.~(\ref{eq:3})--(\ref{eq:4})hold universally, that is they are independent of the form of pair interactions. Eq.~(\ref{eq:4}) was proved in \cite{my3} for the case when the critical point is approached along the line $\lambda_{AA} = 1$. We do not prove it here but one can show that Eq.~(\ref{eq:4}) holds true if the critical point is approached along any line within the stable part of the stability diagram. The argument largely resides on the fact the asymptotic of the binding energy deduced from Eq.~(\ref{eq:5}) dominates over the asymptotic in Eq.~(\ref{eq:3}).

If the 3-body system AAB is bound then from the underlying symmetries  
the ground state wave function can be written either as $\psi(\rho_1, 
\rho_2, \alpha)$ or $\psi(\rho_1, x, \beta)$, where the corner 
angles 
$\alpha , \beta$ are illustrated in Fig.~\ref{Fig:coord} and $\rho_i \equiv |{\pmb \rho}_i|$, $x \equiv |{\pmb x}|$. It is convenient to pass to the units, where $\hbar^{-1}\sqrt{m_A} =1$, in which case ${\pmb \rho}_2 -{\pmb \rho}_1$ equals the Jacobi variable ${\pmb x}$. 
Then one defines the {\bf corner angle distributions} as follows 
\begin{gather}
 F(\alpha) = 8\pi^2 \kappa^3 \sin \alpha \int_0^\infty\int_0^\infty \rho_1^2 \rho_2^2 |\psi(\rho_1, \rho_2 , 
\alpha)|^2 d\rho_1 d\rho_2 , \label{falpha}\\
 \Phi (\beta) = 8\pi^2 \kappa^3 \sin \beta \int_0^\infty\int_0^\infty \rho_1^2 x^2 | \psi(\rho_1, 
x, \beta)|^2 d\rho_1  d x ,  \label{fbeta}
\end{gather}
where we set $\kappa = 2\sqrt{m_B/(m_B + 2m_A)}$. The ground state wave function written in coordinates ${\pmb x},{\pmb y}$ is normalized, as a result for 
corner angle distributions we get 
\begin{equation}
\int_0^\pi F(\alpha) d \alpha =  \int_0^\pi \Phi (\beta) d \beta = 1. 
\end{equation}
(This is the reason for introducing the factor $8\pi^2 \kappa^3$ in Eqs.~(\ref{falpha})-(\ref{fbeta})).

From Eqs.~(\ref{eq:3}), (\ref{eq:4}) we see that the universal behavior in the vicinity of the critical point can be very different. Going around the critical point as shown in Fig.~\ref{Fig:diagram} causes a drastic change in the corner angle distributions. In Fig.~\ref{Fig:diagram} 
we move from the stable point that lies very close to the stability curve above the critical  point to another stable point in the close proximity of the stability curve, which lies below the critical point. At the starting point the distribution has a delta-like shape that concentrates near 
$\alpha= 0$ (for points exactly on the  stability curve that lie above the critical point this distribution makes no sense). At the end point it becomes broad and at the critical point and for all points exactly on the stability curve that lie below the critical point it is well-defined. This bifurcation, which is shown in Fig.~\ref{Fig:bifurcation}, is very similar to a phase transition in statistical physics. The path around the critical point leading to the ``change of phase'' in the sense of corner angle distributions can be arbitrarily short, which is in full analogy with phase transitions around the critical point in thermodynamics. And in full analogy with the physics of phase transitions there is a universal behavior associated with the critical point on the stability diagram.  

\begin{figure}
\includegraphics[height=0.27\textheight]{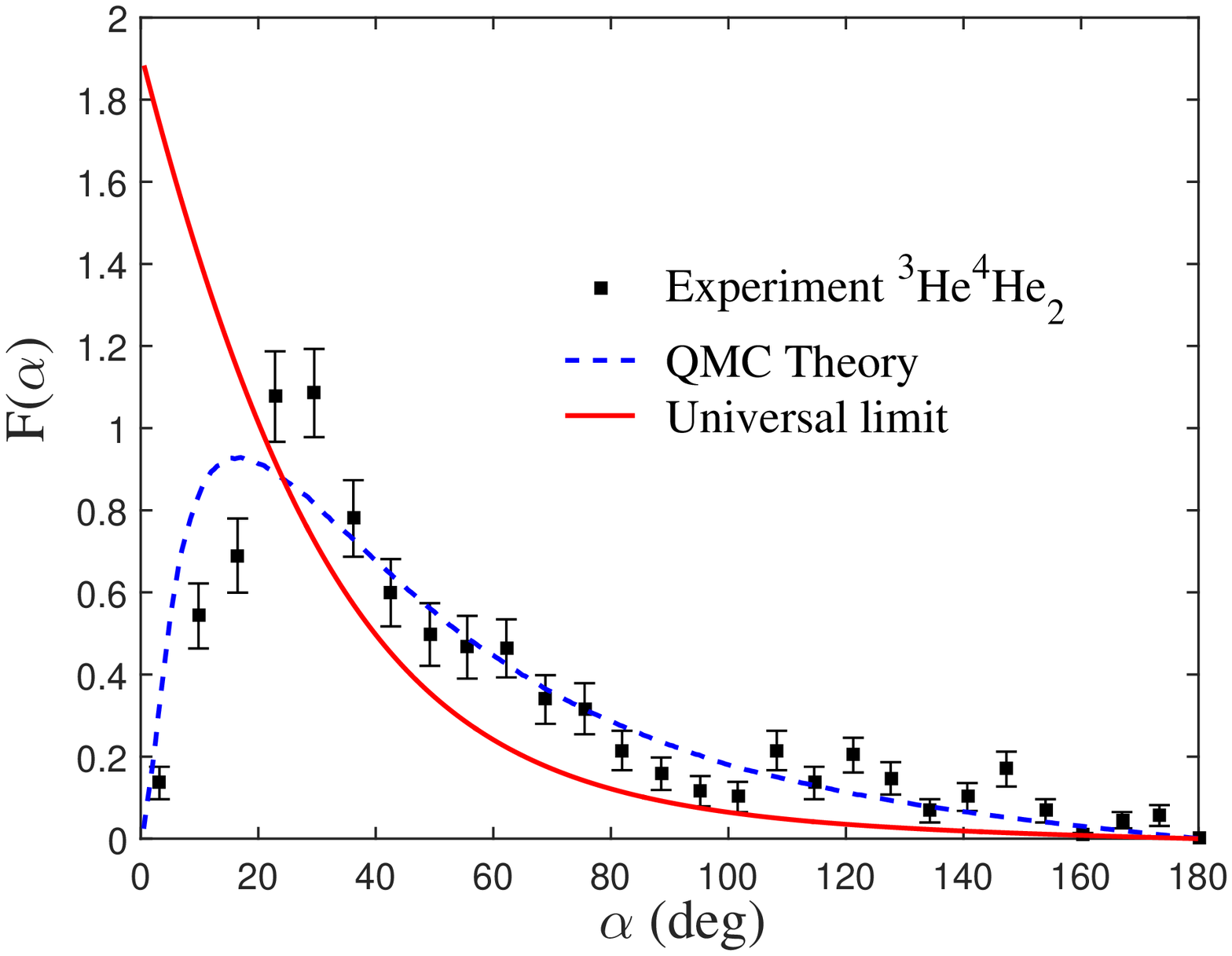}
\includegraphics[height=0.27\textheight]{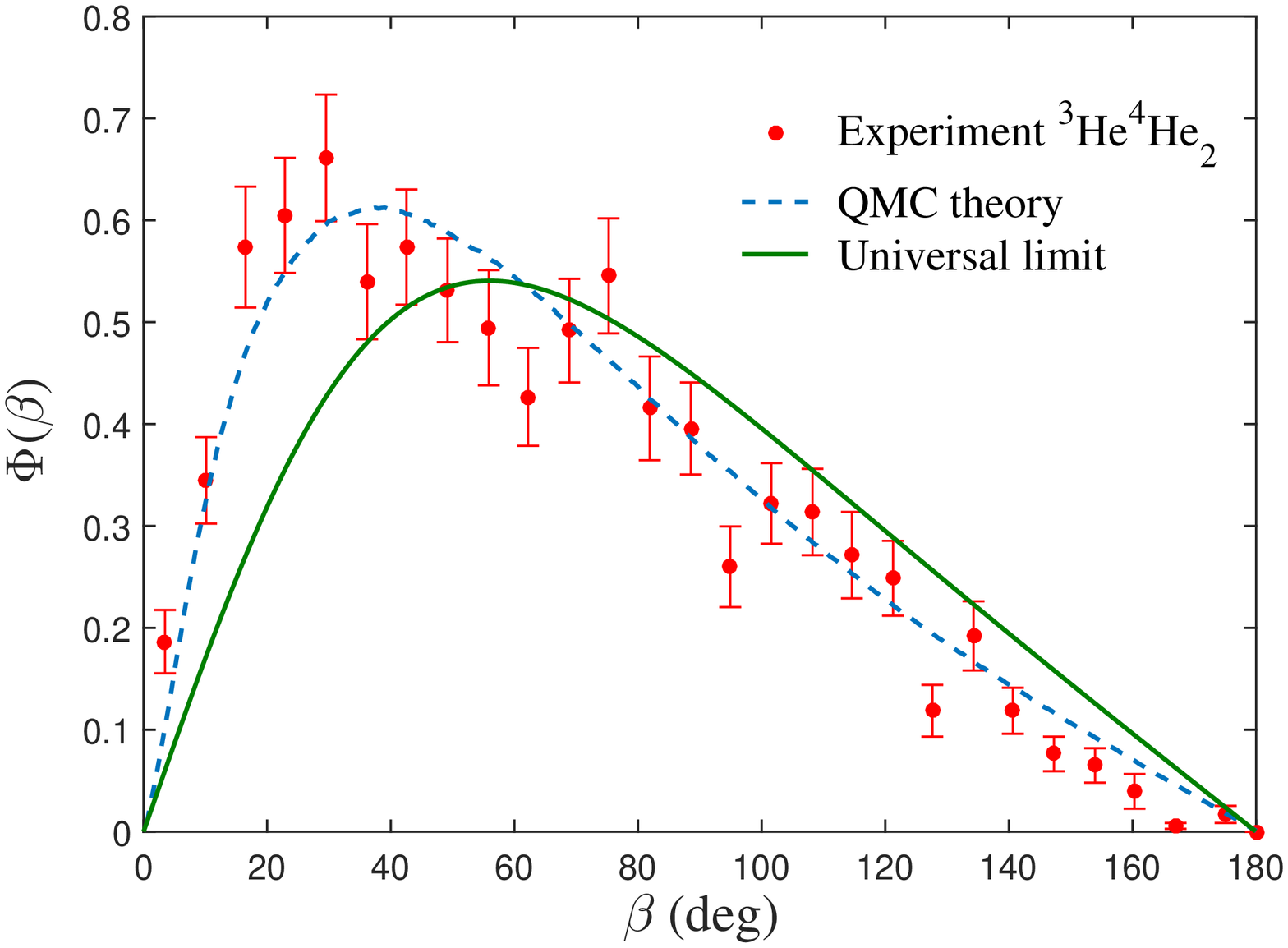}
\caption{\label{Fig:compare1}(Color online). Experimental and theoretical corner angle distributions in the asymmetric Helium trimer compared to the universal limits given by Eqs.~(\ref{fcalpha})-(\ref{fcbeta}), where $m_A/m_B = 4/3$. The experimental data is taken from \cite{doerner2}, the quantum Monte Carlo calculation (QMC) was performed in \cite{bressanini}. The area under all curves is equal to one.}
\end{figure}

If the critical point is approached along the line $\lambda_{AA} = 1$ then 
\begin{gather}
\lim_{\lambda_{AB} \to \lambda_c} \int_0^\pi |F(\alpha) - F_c (\alpha)|d\alpha =0, \label{eq:7}\\
\lim_{\lambda_{AB} \to \lambda_c} \int_0^\pi |\Phi (\beta) - \Phi_c (\beta)|d\beta =0 ,\label{eq:8} 
\end{gather}
where $F_c (\alpha)$ and $\Phi_c (\beta)$ and universal corer angle distributions at the critical point. Eqs. (\ref{eq:7})-(\ref{eq:8}) hold also when the critical point is approached along the line lying in the stable area. In this paper we derive the explicit formulas for the universal corner 
angle distributions
\begin{gather}
 F_c (\alpha) = \frac{4\left(1+\frac{2m_A}{m_B}\right)^{1/2}\sin  \alpha}{\pi(1+\frac{m_A}{m_B})^2} 
I_2\left(\frac{\frac{m_A}{m_B}}{\frac{m_A}{m_B} +1}, \cos \alpha\right) , \label{fcalpha}\\
\Phi_c (\beta) = \frac{\sin \beta}\pi \left(1+\frac{2m_A}{m_B}\right)^{1/2}  I_1\left( \left(\frac 12+\frac{m_A}{2m_B}\right)^{1/2}, \frac{\cos \beta }2 \right) \label{fcbeta}
\end{gather}

Let us introduce the (discontinuous) function $g(x) = \arctan(x)$ for $x 
\geq 0$ and $g(x) = \pi - \arctan(|x|)$ for $x < 0$. 
The following integrals depending on real parameters can be calculated 
analytically
\begin{gather}
 I_1 (s, b) = \int_0^\infty \frac{t^2 dt}{(t^2 -2bt+s^2)^2} 
= \frac{b}{2(s^2-b^2)} + \frac{s^2 g(-\sqrt{s^2-b^2}/b)}{2(s^2-b^2)^{3/2}} 
\label{eq:11}\\
 I_2 (s, b) = \int_0^\infty \frac{t^2 dt}{(1+t^2 -2tb)(t^2 -2bst+1)^2} 
\nonumber\\
= \frac{g(-\frac{\sqrt{1-b^2}}b)}{4b^2\sqrt{1-b^2}(s-1)^2} + \frac{P_1 
(s,b)g(-\frac{\sqrt{1-s^2b^2}}{sb}) + P_2(s,b)}{4b^3(1-s)^4(1-b^2 s^2)^{5/2}} , 
\label{eq:12}
\end{gather}
where 
\begin{gather}
 P_1 (s,b) = (-2bs^2+5sb -4b)(1-s^2b^2)^2 
+ (3b-3bs +bs^2-b^3s)(1-b^2s^2)\nonumber\\
 P_2 (s,b) = (s-3-b^2+3b^2s)(1-s^2b^2)^{3/2} + (3-s)(1-b^2s^2)^{5/2}\nonumber
\end{gather}

The above integrals can be calculated using the following theorem: suppose that 
$\lambda >0$ is non-integer and the polynomial $Q(t)$ is non-zero for $t \geq 0$ 
and 
has degree larger than $1+\lambda$. Then 
\begin{equation}
 \int_0^\infty \frac{t^{\lambda -1}}{Q(t)} dt = \frac{\pi}{\sin (\lambda \pi )} 
\sum_p {\Resi}_p \frac{(-z)^{\lambda -1 }}{Q(z)} , 
\end{equation}
where the sum is over the residues calculated in all poles of $Q(z)$ in the 
complex plain. To calculate the integrals in Eqs.~(\ref{eq:11})-(\ref{eq:12}) we apply his theorem by setting 
$\lambda = 3+\epsilon$ and after that  
letting $\epsilon \to 0$. After the lengthy but straightforward calculation one obtains Eqs.~(\ref{eq:11})-(\ref{eq:12}).

Note one important feature regarding the universal corner angle distribution $F(\alpha)$. For any normalized wave function corresponding to a nonzero binding energy $F(0) =0$. Indeed, the wave function is finite and has an exponential fall off, therefore $F(0) =0$ results from the factor $\sin \alpha$ in Eq.~(\ref{falpha}). At the same time the universal limit for this distribution is such that $F_c (0) \neq 0$, on the contrary, $F_c (\alpha)$ reaches its maximum at zero. $F(\alpha)$ approaches $F_c(\alpha)$ in the sense of Eq.~(\ref{eq:7}), but the convergence in the pointwise sense is nonuniform (this is also discussed in the remark on page 5 in \cite{my3}). In the vicinity of the  critical point $F(\alpha)$ 
starts from zero and goes steeply up staying close to the ordinate axis. Note also that the convergence to the universal limit with the vanishing binding energy is logarithmically slow. This follows from the logarithmic factor in the denominator in Eq.~(\ref{eq:4}), which, in fact, eliminates nonuniversal components in the wave function \cite{my3}.

The asymmetric Helium trimer $^3$He$^4$He$_2$ represents an example of a system that is close to the critical point on the stability diagram. The binding energy of this system is estimated to be 1.23$\times$10$^{-6}$ eV, which is very small on the atomic scale, the pair of atoms $^3$He$_2$ is 
unbound and the Helium dimer $^4$He$_2$ has the binding energy 1.14$\times$10$^{-7}$ eV, which is by an order of magnitude less than that of the trimer.  This data suggests that the asymmetric Helium trimer lies very close to the critical point in the ``tango'' sector on the diagram in Fig.~\ref{Fig:diagram}. The corner angle distributions for the asymmetric trimer were measured experimentally \cite{doerner2}. D. Bressanini \cite{doerner2,bressanini} performed high precision Monte Carlo calculations of the asymmetric trimer using the TTY helium-helium potential \cite{hehepotential}. Fig.~\ref{Fig:compare1} compares the experimental data and the calculations in \cite{bressanini} with the universal limit given by Eqs. (\ref{fcalpha})-(\ref{fcbeta}). Note that the area under all curves is equal to one. The data indicates that the corner angle distributions are relatively close to universal ones.

\begin{figure}
\includegraphics[height=0.35\textheight]{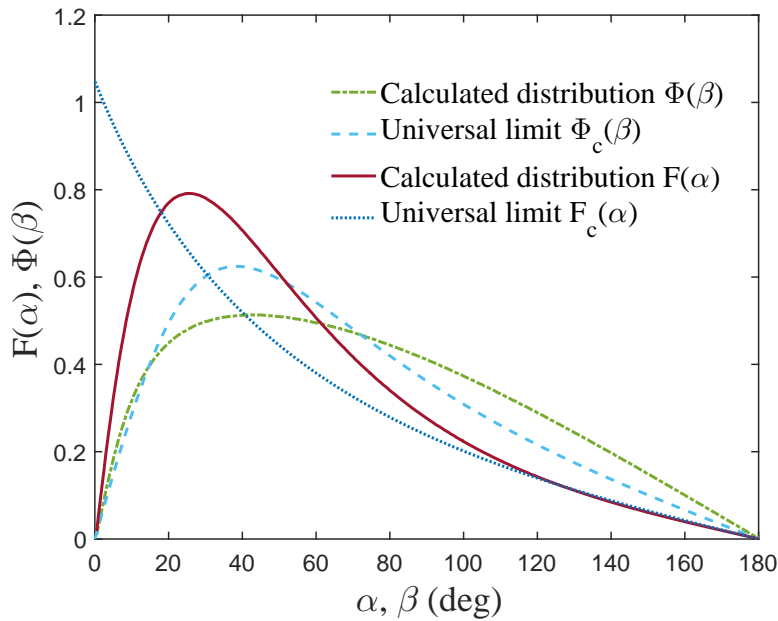}
\caption{\label{Fig:nuclear}(Color online). Calculated corner angle distributions in the nucleus $^{22}$C compared to the universal limits given by Eqs.~(\ref{fcalpha})-(\ref{fcbeta}), where $m_A/m_B = 1/20$. The area under all curves is equal to one.}
\end{figure}

Another example of the system close to the critical point is  the nucleus $^{22}$C, which can be considered as a 3-body system consisting of the core $^{20}$C and two neutrons forming a halo \cite{horiuchi,vaagencarbon}. This system lies in the Borromean square on the diagram in Fig.~\ref{Fig:diagram} because the nucleus $^{21}$C is unbound and two neutrons also do not form a bound pair. The available experimental data indicates \cite{carbonexp} that the nucleus $^{22}$C has an enormously large matter radius, the three-body calculation predicts the biding energy on the order of several KeV (compare this with the binding energy of the deutron 2 MeV, which is considered weakly bound on the nuclear scale). The two neutrons do not form a bound state but have a low lying virtual state at the energy about 143 KeV just above the threshold \cite{tomiocarbon}. This suggests that being viewed as a  3-body system the nucleus $^{22}$C lies in the Borromean sector rather close to the critical point. 

%Two neutrons have a scattering length of 21.X fm, which is X times larger that the effective radius of the nucleon pair interaction. 

We did the QMC calculation of the $^{22}$C nucleus using cluster model used in \cite{vaagencarbon}, which treats this nucleus as a three-body system  consisting of 2 valence neutrons and the core. The neutrons interact with the core through the local potential having the Woods-Saxon shape  \cite{vaagencarbon}
\begin{equation}\label{woods}
V_{cn} (r) = -V_0 \left[ 1+ \exp (a^{-1}(r-R))\right]^{-1}
\end{equation}
We use the same parameters used in \cite{vaagencarbon}, namely, $a = 0.6$ fm and $R = 3.529$ fm. Like in  \cite{vaagencarbon} we tune the depth $V_0$ in order to reproduce the observed matter radius of the nucleus. The neutrons interact through the local Gaussian potential $V_{nn} (r) = -31 \exp(-b^{-2}r^2)$ (MeV), where $b =1.8$ fm. This interaction is tuned \cite{fedorovreports,vaagencarbon,vaagenreports} in order to reproduce the low energy properties of the neutron-neutron system (among them the correct scattering length and no binding in the s channel). Though it is not explicitly mentioned, the neutron-neutron interaction used in \cite{vaagencarbon} is nonlocal, it is set to zero in all  partial waves except the $s$-wave. In the s-wave it is identical with the expression that is used here. Let us remark that for the low binding energy only low energy properties of the interaction matter. In particular, this is illustrated in Table II in \cite{vaagencarbon}, where one can see that the weight of the components in the 3-body wave function for higher partial waves ($l_x \geq 1$) is less than 1\%. Thus our model, where the local neutron-neutron interaction is nonzero in higher partial waves, is very similar to the one used in \cite{vaagencarbon}. The technique of the QMC method is described in \cite{morozi}. 

In Eq.~(\ref{woods}) we set the value $V_0 = 2.84$ MeV and obtain the separation energy of two neutrons $S_{2n} = 0.0052(4)$ MeV and the mean squared hyperradius $\langle \rho^2 \rangle = 250(12)$ fm$^2$. According to Eq.~(2) in \cite{vaagencarbon} this gives $4.4$ fm for the root mean square matter radius, which is in accordance with the results listed in the Table I in \cite{vaagencarbon}. In spite of very small binding energy this value is less than 
the mean value and nearly equals the lower bound for the experimental matter radius \cite{carbonexp}. Fig.~\ref{Fig:nuclear} compares the corner angle distribution for the calculated three-body wave function with the universal limits given by Eqs.~(\ref{fcalpha})-(\ref{fcbeta}). The deviation from 
the universal limit results from the finite scattering length in the neutron-neutron interaction.

Let us pass to the derivation of Eqs.~(\ref{fcalpha})-(\ref{fcbeta}). From Eqs. (3), (5) in \cite{my3} we know that when the critical point is approached 
\begin{equation}\label{key}
\int \rho^5 |\psi (\rho, \theta, \hat {\pmb x}, \hat {\pmb y})|^2 d\rho \to \frac 1{4\pi^3 \cos^2 \theta} , 
\end{equation}
where $\rho^2 = {\pmb x}^2 + {\pmb y}^2 $, $\tan \theta = |{\pmb y}|/|{\pmb x}|$ and $\hat {\pmb x} , \hat {\pmb y}$ are unit vectors in the directions of ${\pmb x}, {\pmb y}$ respectively. Eq.~(\ref{key}) is the key point in the derivation. Note that with the chosen scales ${\pmb y} = \kappa \tilde {\pmb y}$, where  the vector $\tilde {\pmb y}$ is pictured in Fig.~\ref{Fig:coord}. In Eq.~(\ref{fbeta}) let us change the integration variables from $x, \rho_1$ to $\rho , t$, where $t = \rho_1 /x$. By a geometric argument (see Fig.~\ref{Fig:coord} ) we get
\begin{equation}\label{pmby}
\rho^2 = {\pmb x}^2 + \kappa^{2} {\tilde {\pmb y}}^2 = \left(1 + \frac{\kappa^2}4\right) x^2 + \kappa^2 \rho_1^2 - \kappa^2 x\rho_1 \cos \beta 
\end{equation}
Then the Jacobian of the transformation reads
\begin{gather}
J(\rho, t) = \begin{vmatrix}
\frac{\partial \rho}{\partial x}& \frac{\partial \rho}{\partial \rho_1}\\
\frac{\partial t}{\partial x}& \frac{\partial t}{\partial \rho_1}
\end{vmatrix}^{-1} = \rho\left[ (1+\kappa^2/4) -\kappa^2t\cos\beta +\kappa^2 t^2 \right]^{-1} \label{jacobian}
\end{gather}
We obviously have 
\begin{equation}\label{product}
x^2\rho_1^2 = t^2 x^4 = t^2 \rho^4 \cos^4 \theta
\end{equation}
Thereby by Eq.~(\ref{pmby})
\begin{equation}\label{cos2}
\cos^2 \theta = \frac{x^2}{\rho^2} = \frac 1{(1+\kappa^2/4) -\kappa^2t\cos\beta +\kappa^2 t^2}
\end{equation}
Substituting Eq.~(\ref{product}) and Eq.~(\ref{jacobian}) into Eq.~(\ref{fbeta}) we obtain 
\begin{gather}
 \Phi (\beta) = 8\pi^2 \kappa^3 \sin \beta \int_0^\infty \int_0^\infty t^2 \rho^4 \cos^4 \theta |\psi(x (\rho, t), \rho_1 (\rho, t) , \beta)|^2 J(\rho, t)d\rho  dt \\
 = 8\pi^2 \kappa^3 \sin \beta \int_0^\infty dt t^2 \cos^6 \theta \int_0^\infty  \rho^5 |\psi(x (\rho, t), \rho_1 (\rho, t) , \beta)|^2 d\rho  .  
\end{gather}
Now we use Eq.~(\ref{cos2}) and Eq.~(\ref{key}) to get the expression for the critical corner angle distribution
\begin{equation}
 \Phi_c (\beta) = \frac{2  \kappa^3\sin \beta}{\pi}  \int_0^\infty t^2 \cos^4 \theta dt = \frac{2 \kappa^3 \sin \beta}{\pi}  \int_0^\infty \frac{t^2 dt}{[(1+\kappa^2/4) -\kappa^2t\cos\beta +\kappa^2 t^2 ]^2} 
\end{equation}
Substituting the expression for $\kappa$ in terms of masses we get Eq.~(\ref{fcbeta}). Now let us prove Eq.~(\ref{fcalpha}). In Eq.~(\ref{falpha}) let us change the integration variables from $\rho_1, \rho_2$ to $\rho , t'$, where $t' = \rho_2 /\rho_1$. 
From the geometry in Fig.~\ref{Fig:coord} 
\begin{equation}
\rho^2 = {\pmb x}^2 + \kappa^{2} {\tilde {\pmb y}}^2 =  (1+\kappa^2/4)(\rho_1^2 + \rho_2^2) -2\rho_1 \rho_2 (1-\kappa^2/4)\cos \alpha \label{rhosc}
\end{equation} 
For convenience let us introduce 
\begin{equation}\label{xi}
\xi (t', \alpha) = \left[ (1+\kappa^2/4)(1+t'^2) -2(1-\kappa^2/4) t' \cos\alpha \right]^{-1}
\end{equation}

Now we can calculate the Jacobian of the transformation 
\begin{gather}
J(\rho, t') = \begin{vmatrix}
\frac{\partial \rho}{\partial \rho_1}& \frac{\partial \rho}{\partial \rho_2}\\
\frac{\partial t'}{\partial \rho_1}& \frac{\partial t'}{\partial \rho_2}
\end{vmatrix}^{-1} = \rho \xi (t', \alpha) . 
\end{gather}
Using Eq.~(\ref{rhosc}) we also get 
\begin{equation}
\rho_1^2 \rho_2^2 = \rho_1^4 t'^2 = t'^2 \rho^4 \left[ \xi (t', \alpha) \right]^{2} . 
\end{equation}
Eq.~(\ref{cos2}) can be rewritten as follows
\begin{equation}\label{cos22}
\cos^2 \theta = \frac{x^2}{\rho^2} = (1+t'^2 -2t'\cos\alpha )\xi (t', \alpha) . 
\end{equation}
Passing to the new integration variables we get from Eq.~(\ref{falpha}) 
\begin{gather}
F (\alpha) = 8\pi^2 \kappa^3 \sin \alpha \int_0^\infty dt' \left[ \xi (t', \alpha) \right]^{3} t'^2 \int_0^\infty  \rho^5 |\psi(\rho_1 (\rho, t'), \rho_2 (\rho, t') , \alpha)|^2 d\rho 
\end{gather}
Hence, due to Eq.~(\ref{cos22}) and Eq.~(\ref{key}) 
\begin{gather}
F_c (\alpha) = \frac{2  \kappa^3\sin \alpha}{\pi}   \int_0^\infty \frac{\left[ \xi (t', \alpha) \right]^{2} t'^2}{1+t'^2 -2t'\cos\alpha } dt' , 
\end{gather}
where one should substitute the explicit expression for $ \xi (t', \alpha) $ given by Eq.~(\ref{xi}). After the substitution of $\kappa$ in terms of masses we get Eq.~(\ref{fcalpha}). 

We thank S. Ershov, J. S. Vaagen, M. Zhukov, M. Kunitski and R. D\"orner for stimulating discussions. One of us (D.G.) is grateful to Horst St\"ocker for supporting the project at FIAS.

\end{document}